\documentclass[runningheads]{llncs}

\usepackage{graphicx}
\usepackage{amsmath}
\usepackage{amssymb}
\usepackage{booktabs}
\usepackage{xcolor}
\usepackage{url}
\usepackage{cite}

\begin{document}

\title{Memetic Search for Supersingular Elliptic Curves over $\mathbb{F}_p$}

\titlerunning{Memetic Search for Supersingular Curves over $\mathbb{F}_p$}

\author{
Ismel Martínez-Díaz\inst{1}\orcidID{0000-0002-7064-232X}
}

\authorrunning{I. Martínez-Díaz}

\institute{
Department of Mathematics, Universitat de Lleida,\\
C/ Jaume II, 69, 25001 Lleida, Spain\\
\email{imd6@alumnes.udl.cat}
}

\maketitle

\begin{abstract}

The search for supersingular elliptic curves is a fundamental computational
problem in isogeny-based cryptography. A recent metaheuristic formulation over
$\mathbb{F}_{p^2}$ introduced the NonMultiplicity Distance (NMD) objective,
measuring the deviation of the Frobenius trace from a multiple of $p$, and
showed that uninformed random search fails beyond $\approx 10^{13}$ candidates.
This work investigates metaheuristic search over the prime field $\mathbb{F}_p$,
the setting for oriented isogeny protocols such as CSIDH, OSIDH, and SQISign.
Although the candidate space decreases from $p^2$ to $p$, the supersingular locus
is asymptotically sparse ($O(\sqrt{p}\log p)$ curves), keeping the search
exponentially difficult. We formulate a memetic algorithm tailored to
$\mathbb{F}_p$ using a one-dimensional $j$-invariant chromosome, bit-level
recombination, adaptive mutation, and periodic local search under the NMD objective.
Benchmarks across 30 independent seeds at 40-bit, 46-bit, and 51-bit prime sizes
($p \approx 1.13\times 10^{15}$) show that the algorithm discovers an exact
supersingular curve at 46 bits and consistently converges to ``near-supersingular''
ordinary curves with Frobenius traces remarkably close to zero: best NMD values of
19 at 40 bits and 3 at 51 bits, corresponding to relative trace deviations of
$1.3\times 10^{-5}$ and $4.5\times 10^{-8}$ across the Hasse interval. These
results demonstrate that NMD-driven memetic search effectively navigates the
sparse $\mathbb{F}_p$ landscape and systematically locates near-supersingular
structures.


\end{abstract}

\section{Introduction}
\label{sec:intro}

Supersingular elliptic curves sit at the intersection of classical number
theory and modern post-quantum cryptography. This section traces the
historical path that made them central to isogeny-based protocols, reviews
the computational challenge of generating them, explains why the present
work moves the metaheuristic search from $\mathbb{F}_{p^2}$ to the prime
field $\mathbb{F}_p$, and summarizes our contributions and the organization
of the paper.

\subsection{Historical Context and the Post-Quantum Paradigm}

Elliptic curves have served as foundational building blocks of modern public-key
cryptography since their independent introduction by Miller~\cite{miller1985}
and Koblitz~\cite{koblitz1987} in the mid-1980s. For decades, classical
elliptic curve cryptography (ECC) has relied on the intractability of the
Elliptic Curve Discrete Logarithm Problem (ECDLP) in the group of rational
points $E(\mathbb{F}_q)$, supported by efficient point-counting algorithms
pioneered by Schoof~\cite{schoof1985} and subsequent Elkies-Atkin improvements.
Pairing-based cryptography later expanded this domain through the Weil and
Tate pairings on supersingular and pairing-friendly curves~\cite{boneh2001,galbraith2001}.

However, the advent of quantum computation poses an existential threat to
classical public-key systems: Shor's algorithm~\cite{shor1994} solves both
integer factorization and discrete logarithms in polynomial time, rendering
traditional RSA, Diffie-Hellman, and ECDLP-based schemes vulnerable. In
response, post-quantum cryptography (PQC) has emerged to design cryptosystems
resistant to quantum cryptanalysis.

Among post-quantum candidates, \emph{isogeny-based cryptography} offers an
appealing framework characterized by compact key sizes and rich underlying
algebraic structures. The foundational ideas trace back to Couveignes'
hard homogeneous spaces~\cite{couveignes2006} and Rostovtsev and
Stolbunov's public-key schemes~\cite{rostovtsev2006}, which used group actions
on ordinary elliptic curves. The field shifted dramatically when Jao and
De Feo~\cite{jao2011} introduced the Supersingular Isogeny Diffie-Hellman
(SIDH) protocol, transitioning from ordinary curves to supersingular curves
defined over $\mathbb{F}_{p^2}$ to exploit the non-commutative structure and
optimal Ramanujan expansion properties of supersingular isogeny graphs.

In 2022, devastating polynomial-time attacks utilizing auxiliary torsion-point
information, discovered by Castryck and Decru~\cite{castryck2022} and Maino
and Martindale~\cite{maino2022}, broke SIDH and its standardized variant SIKE.
Crucially, these attacks specifically exploited the auxiliary point
evaluations shared during SIDH key exchange, leaving the fundamental problem
of computing isogenies between supersingular curves uncompromised.

Consequently, modern post-quantum research has pivoted toward new paradigms
that avoid auxiliary point disclosure. Chief among these are:
\begin{enumerate}
   \item \textbf{Group-action protocols over prime fields}: Commutative
         isogeny schemes such as CSIDH~\cite{castryck2018}, OSIDH~\cite{colo2020},
         and SCALLOP~\cite{defeo2023scallop}, which operate directly on the
         set of $\mathbb{F}_p$-rational supersingular elliptic curves under the
         action of the ideal class group $\operatorname{cl}(\mathcal{O})$.
   \item \textbf{Oriented supersingular curves and signatures}: Modern
         constructions such as SQISign~\cite{defeo2020sqisign} and general
         oriented isogeny protocols~\cite{arpin2024}, which rely on curves
         equipped with explicit subring embeddings (orientations) in their
         endomorphism rings.
\end{enumerate}
In both settings, supersingular elliptic curves defined over the prime field
$\mathbb{F}_p$ occupy a central role.

\subsection{The Challenge of Supersingular Curve Generation}

Constructing supersingular elliptic curves is therefore a core problem with
immediate cryptographic utility. Deterministic methods exist: Bröker~\cite{broker2009}
proposed a polynomial-time algorithm using class field theory and complex
multiplication (CM), and Sutherland~\cite{sutherland2012} developed Hilbert
class polynomial and modular polynomial techniques. However, deterministic CM
constructions require computing class polynomials whose degrees grow with the
class number $h(-D)$, necessitating significant algebraic machinery.

Treating curve discovery as an optimization problem offers an alternative
metaheuristic approach. Martínez-Díaz et al.~\cite{martinez2025} recently
formulated the search for supersingular curves over $\mathbb{F}_{p^2}$ as a
combinatorial optimization task. They introduced the NonMultiplicity
Distance (NMD) as an objective function based on the Frobenius trace and
demonstrated that evolutionary and local search heuristics could exploit the
resulting fitness landscape. However, their empirical findings revealed a
clear scalability threshold: once the candidate space exceeded approximately
$10^{13}$ curves, naive random search became ineffective, and metaheuristic
exploration encountered severe stagnation. The same study also observed that
the NMD objective tends to drive the search towards \emph{ordinary} curves
whose Frobenius trace is very close to zero, and suggested that such
``near-supersingular'' curves deserve study in their own right.

\subsection{Motivation and Shift to the Prime Field $\mathbb{F}_p$}

The present work reformulates the metaheuristic search from $\mathbb{F}_{p^2}$
to the prime field $\mathbb{F}_p$. Why shift the search space?
\begin{itemize}
   \item \textbf{Combinatorial Challenge vs. Search Complexity}: Over
         $\mathbb{F}_{p^2}$, the total space of $j$-invariants is $p^2$, and
         there are $\approx p/12$ supersingular $j$-invariants, yielding a
         density of $\approx 1/(12p)$. Over $\mathbb{F}_p$, the total candidate
         space is reduced to $p$. However, the number of supersingular
         $j$-invariants in $\mathbb{F}_p$ is only $O(\sqrt{p}\log p)$~\cite{delfs2016}.
         The resulting hit probability $O(\log p/\sqrt{p})$ remains exponentially
         small in $\log p$. Thus, for cryptographically relevant primes ($p \ge 10^{13}$),
         finding a supersingular curve in $\mathbb{F}_p$ remains a formidable
         needle-in-a-haystack problem where uninformed search inevitably fails.
   \item \textbf{A One-Dimensional Search Space}: Over $\mathbb{F}_p$ the
         chromosome is a single field element $j \in \mathbb{F}_p$, which
         admits natural integer-arithmetic neighborhoods, bit-level
         recombination, and a direct, unambiguous notion of ``distance'' on
         which mutation radii and local search can be scheduled. This removes
         the extension-field representation overhead of the
         $\mathbb{F}_{p^2}$ formulation.
   \item \textbf{Direct Applicability to Oriented Protocols}:
         $\mathbb{F}_p$-rational supersingular curves naturally possess an
         orientation by the imaginary quadratic order $\mathbb{Z}[\pi] \cong \mathbb{Z}[\sqrt{-p}]$
         given by the Frobenius endomorphism $\pi$. Finding curves directly in
         $\mathbb{F}_p$ yields the precise starting curves required by modern
         oriented protocols (CSIDH, OSIDH, SCALLOP, and oriented signature schemes),
         which cannot directly utilize general $\mathbb{F}_{p^2}$ supersingular curves.
\end{itemize}

\subsection{Contributions and Paper Organization}

This paper makes the following main contributions:
\begin{enumerate}
   \item \textbf{Formulation of $\mathbb{F}_p$ Memetic Search}: We adapt the
         combinatorial optimization model of Martínez-Díaz et al.~\cite{martinez2025}
         to the prime field $\mathbb{F}_p$, designing a one-dimensional chromosome
         representation, bit-level crossover operators, adaptive mutation
         radius schedules, and local search routines, while keeping the NMD
         objective function unchanged.
   \item \textbf{Experimental Validation up to $10^{15}$ Scale}: We evaluate
         random search and the memetic algorithm on three primes of increasing
         size -- 40-bit, 46-bit, and a 51-bit challenge prime
         ($p \approx 1.13 \times 10^{15}$) -- over 30 independent seeds per
         configuration. The memetic algorithm reaches an exact, independently
         verified supersingular curve at the 46-bit benchmark, and at the
         other two scales converges to ordinary curves with Frobenius trace
         $|t| = 19$ and $|t| = 3$, respectively.
   \item \textbf{Near-Supersingular Ordinary Curves}: We quantify how close
         the memetic search gets to supersingularity relative to the Hasse
         interval, show that this behavior is systematic rather than
         accidental, and argue -- reinforcing the observation of
         Martínez-Díaz et al.~\cite{martinez2025} -- that NMD-driven search over
         $\mathbb{F}_p$ is an efficient generator of near-supersingular curves
         worthy of independent study.
\end{enumerate}

The remainder of this paper is organized as follows: Section~\ref{sec:background}
provides mathematical background on supersingular curves and memetic algorithms.
Section~\ref{sec:search-space} analyzes the search spaces and problem
formulation. Section~\ref{sec:objective} describes the NMD objective function.
Section~\ref{sec:memetic} details the proposed memetic algorithm.
Section~\ref{sec:experiments} presents the experimental setup and empirical
results. Section~\ref{sec:discussion} discusses implications, and
Section~\ref{sec:conclusion} concludes with future research directions.

\section{Background and Preliminaries}
\label{sec:background}

This section collects the mathematical and algorithmic notions on which the
rest of the paper depends. We first recall elliptic curves over prime fields
and the role of the Frobenius trace in characterizing supersingularity, and
then summarize the memetic algorithm paradigm that drives our search.

\subsection{Elliptic Curves and the Frobenius Trace}

Let $p > 3$ be prime, and let $\mathbb{F}_p$ denote the finite field of $p$
elements. An elliptic curve $E$ defined over $\mathbb{F}_p$ can be described
by the short Weierstrass equation:
\begin{equation}
   E/\mathbb{F}_p: \quad y^2 = x^3 + Ax + B, \qquad A, B \in \mathbb{F}_p,
\end{equation}
with non-zero discriminant $\Delta = -16(4A^3 + 27B^2) \neq 0$.

The $j$-invariant of $E$, given by
\begin{equation}
   j(E) = 1728 \frac{4A^3}{4A^3 + 27B^2},
\end{equation}
classifies $E$ up to isomorphism over the algebraic closure $\overline{\mathbb{F}}_p$.
For any $j \in \mathbb{F}_p \setminus \{0, 1728\}$, a canonical curve representative
having $j(E_j) = j$ is given by:
\begin{equation}
   E_j: \quad y^2 = x^3 + \frac{3j}{1728-j}x + \frac{2j}{1728-j}.
   \label{eq:canonical-curve}
\end{equation}
The boundary cases $j=0$ and $j=1728$ correspond respectively to curves of the
form $y^2 = x^3 + B$ and $y^2 = x^3 + Ax$.

The $p$-power Frobenius endomorphism $\pi: E \to E$ is defined by
$\pi(x,y) = (x^p, y^p)$. The number of $\mathbb{F}_p$-rational points is
governed by Hasse's theorem:
\begin{equation}
   \#E(\mathbb{F}_p) = p + 1 - t,
\end{equation}
where the Frobenius trace $t \in \mathbb{Z}$ satisfies the Hasse bound
$|t| \leq 2\sqrt{p}$.

\begin{definition}[Supersingularity]
An elliptic curve $E/\mathbb{F}_p$ is \textbf{supersingular} if and only if
$t \equiv 0 \pmod p$. For primes $p > 3$, the Hasse bound $|t| \leq 2\sqrt{p} < p$
implies that $E/\mathbb{F}_p$ is supersingular if and only if:
\begin{equation}
   t = 0 \quad \Longleftrightarrow \quad \#E(\mathbb{F}_p) = p + 1.
\end{equation}
\end{definition}

For supersingular curves, the full endomorphism ring $\operatorname{End}(E)$
over $\overline{\mathbb{F}}_p$ is a maximal order in a definite quaternion algebra
$B_{p,\infty}$ ramified at $p$ and $\infty$~\cite{galbraith2012}. In contrast,
for ordinary curves, $\operatorname{End}(E)$ is an order in the imaginary
quadratic field $\mathbb{Q}(\sqrt{t^2 - 4p})$. Over the base field
$\mathbb{F}_p$, the $\mathbb{F}_p$-rational endomorphism ring of a supersingular
curve is an order in $\mathbb{Q}(\sqrt{-p})$.

\subsection{Memetic Algorithms}

Memetic Algorithms (MAs)~\cite{moscato1989} are population-based metaheuristics
that synergistically combine global evolutionary exploration with problem-specific
local refinement (local search / meme exploitation). The evolutionary operators
(selection, recombination, and mutation) preserve population diversity across
the global search space, while local search intensifies exploitation around
promising candidate regions.

In combinatorial and arithmetic domains where the objective function presents
complex, non-linear, or deceptive landscapes, memetic algorithms frequently
outperform pure genetic algorithms and random sampling by avoiding premature
convergence and navigating rugged fitness basins~\cite{wolpert1997}.

\section{Search Space and Problem Formulation}
\label{sec:search-space}

Before describing the algorithm, we make precise the combinatorial object
being searched. This section contrasts the $\mathbb{F}_{p^2}$ search space
of previous work with the $\mathbb{F}_p$ space adopted here, quantifies the
density of supersingular $j$-invariants in each, and motivates the concrete
prime sizes used as experimental benchmarks.

\subsection{Comparison: Search over $\mathbb{F}_{p^2}$ vs. $\mathbb{F}_p$}

The choice of search domain fundamentally determines the combinatorial and
arithmetic characteristics of the optimization problem. Table~\ref{tab:search-space}
summarizes the key differences between searching over $\mathbb{F}_{p^2}$
(the setting of Martínez-Díaz et al.~\cite{martinez2025}) and over $\mathbb{F}_p$
(the present work).

\begin{table}[htbp]
\centering
\caption{Comparison of the search spaces and density characteristics.}
\label{tab:search-space}
\setlength{\tabcolsep}{5pt}
\begin{tabular}{lcccc}
\toprule
\textbf{Field} & \textbf{Space Size} & \textbf{Supersingular Count} & \textbf{Density} & \textbf{Random Hit Scale} \\
\midrule
$\mathbb{F}_{p^2}$ & $p^2$ & $\approx p/12$ & $\approx \frac{1}{12p}$ & $O(p)$ \\
$\mathbb{F}_p$ & $p$ & $O(\sqrt{p}\log p)$ & $O\left(\frac{\log p}{\sqrt{p}}\right)$ & $O\left(\frac{\sqrt{p}}{\log p}\right)$ \\
\bottomrule
\end{tabular}
\end{table}

Over $\mathbb{F}_{p^2}$, the candidate space has size $p^2$, and the number of
supersingular $j$-invariants is $\lfloor p/12 \rfloor + \varepsilon$ (where $\varepsilon \in \{0,1,2\}$).
Over $\mathbb{F}_p$, the candidate space is $p$, and the number of $\mathbb{F}_p$-rational
supersingular $j$-invariants is bounded by $O(\sqrt{p}\log p)$ as established by
Delfs and Galbraith~\cite{delfs2016}.

Restricting the search to $\mathbb{F}_p$ offers two structural advantages:
\begin{enumerate}
   \item \textbf{Dimensionality reduction}: The optimization space is
         one-dimensional ($j \in \mathbb{F}_p$), eliminating complex
         extension-field representation overhead and giving mutation and
         local search a natural integer metric to operate on.
   \item \textbf{Cryptographic alignment with oriented protocols}: Modern
         oriented isogeny schemes (CSIDH~\cite{castryck2018}, OSIDH~\cite{colo2020},
         SCALLOP~\cite{defeo2023scallop}) explicitly require $\mathbb{F}_p$-rational
         supersingular curves as public starting curves to instantiate the class
         group action $\operatorname{cl}(\mathbb{Z}[\sqrt{-p}]) \star E$.
\end{enumerate}

Despite the higher relative density in $\mathbb{F}_p$ compared to $\mathbb{F}_{p^2}$,
the probability $O(\log p/\sqrt{p})$ remains vanishingly small for large primes:
for a 46-bit prime ($p \approx 3.5 \times 10^{13}$), the density is on the order of
$10^{-6}$, meaning uninformed random search requires millions of point counts
and, within feasible evaluation budgets, succeeds only by chance.

\subsection{Experimental Benchmark Selection}

Martínez-Díaz et al.~\cite{martinez2025} observed that naive random search
stagnated when the search space exceeded approximately $10^{13}$ candidates
($2^{43} \lesssim N \lesssim 2^{47}$). For our $\mathbb{F}_p$ formulation,
the candidate cardinality is $N = p$. Following the same escalating
validation strategy used by Martínez-Díaz et al. to probe scalability, we
add a third, more challenging prime beyond their $10^{13}$--$10^{14}$ regime.

We adopt a three-stage benchmarking methodology:
\begin{table}[htbp]
\centering
\caption{Selected prime parameters for the experimental study.}
\label{tab:selected-primes}
\setlength{\tabcolsep}{3pt}
\small
\begin{tabular}{lcccc}
\toprule
\textbf{Stage} & \textbf{Bits} & \textbf{Prime $p$} & \textbf{Space $|\mathcal{S}_p|$} & \textbf{Purpose} \\
\midrule
Calibration & 40 & $549{,}755{,}813{,}927$ & $5.50 \times 10^{11}$ & Calibration \& baseline \\
Benchmark   & 46 & $35{,}184{,}372{,}088{,}907$ & $3.52 \times 10^{13}$ & $10^{13}$-scale benchmark \\
Ext.\ Challenge & 51 & $1{,}125{,}899{,}906{,}842{,}679$ & $1.13 \times 10^{15}$ & Beyond-$10^{13}$ scaling \\
\bottomrule
\end{tabular}
\end{table}

The 46-bit prime places our search space directly into the $3.52 \times 10^{13}$
candidate scale used as the primary benchmark, while the 51-bit prime
($p \approx 1.13 \times 10^{15}$, about $32\times$ larger) tests how the
memetic search degrades well beyond the regime originally identified as
problematic. All three primes satisfy $p \equiv 3 \pmod 4$, so that
$y^2 = x^3 + x$ ($j = 1728$) is supersingular and the boundary case is
handled uniformly by the implementation.

\section{Objective Function}
\label{sec:objective}

We retain the NonMultiplicity Distance (NMD) objective function introduced by
Martínez-Díaz et al.~\cite{martinez2025} without modification, ensuring that
the optimization target is preserved while changing the search space.

For an elliptic curve $E/\mathbb{F}_p$, let $t = p + 1 - \#E(\mathbb{F}_p)$ be
its Frobenius trace. The NonMultiplicity Distance is defined as:
\begin{equation}
   \operatorname{NMD}(E) = \min\Big( t \bmod p, \; p - (t \bmod p) \Big).
   \label{eq:nmd}
\end{equation}
The minimization problem is formulated as:
\begin{equation}
   \boxed{ \min_{j \in \mathbb{F}_p} f(j) = \operatorname{NMD}(E_j) }
\end{equation}
where $E_j$ is the canonical curve representative defined in~(\ref{eq:canonical-curve}),
satisfying:
\begin{equation}
   0 \le f(j) \le \left\lfloor \frac{p}{2} \right\rfloor.
\end{equation}
A fitness value $f(j) = 0$ indicates that $t \equiv 0 \pmod p$. For primes $p > 3$,
this is both necessary and sufficient for $E_j$ to be supersingular.

Because $|t| \le 2\sqrt{p} < p/2$ for all $p > 16$, the NMD of any curve over
$\mathbb{F}_p$ is simply $|t|$. The objective is therefore a direct measure
of the absolute Frobenius trace, and its natural scale is the width
$2\sqrt{p}$ of the Hasse interval rather than $p$ itself. We use this
observation in Section~\ref{sec:experiments} to interpret non-zero NMD values.

\section{Proposed Memetic Approach}
\label{sec:memetic}

The proposed method is a memetic algorithm whose population lives directly in
$\mathbb{F}_p$. We describe the one-dimensional chromosome encoding and its
evaluation via point counting, followed by the evolutionary operators --
bit-wise recombination, adaptive mutation, and periodic local search -- that
have been adapted to this finite-field representation.

\subsection{Chromosome Representation and Evaluation}

An individual chromosome is encoded as a single field element:
\begin{equation}
   \chi = j \in \{0, 1, \dots, p-1\}.
\end{equation}
Evaluating an individual consists of constructing the canonical curve $E_j$
via~(\ref{eq:canonical-curve}), computing the curve order $\#E_j(\mathbb{F}_p)$
via SEA/point-counting, determining the trace $t$, and evaluating $f(j) = \operatorname{NMD}(E_j)$.
The initial population is sampled uniformly at random from $\mathbb{F}_p$
without repetition.

\subsection{Evolutionary Operators}

\subsubsection{Bit-Wise Recombination (Crossover).}
Given two parent chromosomes $j_1, j_2 \in \mathbb{F}_p$, we represent them in
binary form:
\begin{equation}
   j_1 = \sum_{k=0}^{n-1} b_{1,k} 2^k, \qquad j_2 = \sum_{k=0}^{n-1} b_{2,k} 2^k, \qquad n = \lceil \log_2 p \rceil.
\end{equation}
A bit interval $[\kappa_1, \kappa_2]$ is chosen uniformly at random, and bits within
the interval are swapped between parents. The resulting integer is reduced modulo $p$.
This enables recombination over single finite field elements.

\subsubsection{Adaptive Neighborhood Mutation.}
For an individual $j$, mutation produces an offspring:
\begin{equation}
   j' = (j + \delta) \bmod p, \qquad \delta \in_R [-r_g, r_g],
\end{equation}
where the mutation radius $r_g$ decays linearly across generations
$g \in \{0, \dots, G_{\max}-1\}$:
\begin{equation}
   r_g = r_{\text{init}} \left(1 - \frac{g}{G_{\max}-1}\right)
       + r_{\text{final}} \frac{g}{G_{\max}-1}.
\end{equation}
This schedule begins with broad global exploration ($r_{\text{init}} = \lfloor p/4 \rfloor$)
and anneals toward fine-grained local refinement ($r_{\text{final}} = \lfloor p/2^{16} \rfloor$).

\subsubsection{Selection and Elitism.}
Parents are chosen by tournament selection of size three. The best
$N_{\text{elite}}$ individuals of each generation are carried over unchanged.

\subsubsection{Local Search Exploitation.}
Periodically (every 10 generations), an intensification phase is triggered on
the elite individuals. For an elite candidate $j$, local search explores an
integer neighborhood $N_\rho(j) = \{(j + k) \bmod p : -\rho \le k \le \rho\}$
with a radius $\rho$ that is halved after each round of unsuccessful
attempts, greedily accepting improvements $f(j') < f(j)$.

\section{Experimental Evaluation and Results}
\label{sec:experiments}

We now evaluate uniform random sampling and the memetic algorithm on the
primes of Table~\ref{tab:selected-primes}. After fixing the experimental
configuration, we compare success rates and best fitness across all
configurations, independently verify the supersingular curve found, and
examine the convergence behavior of the memetic search and the
near-supersingular curves it returns.

\subsection{Experimental Configuration}

All algorithms were implemented and evaluated in SageMath 10.x. Experiments
were executed across 30 independent random seeds per configuration --
a standard sample size for statistically meaningful success-rate
estimates -- for each of the three primes in Table~\ref{tab:selected-primes}:
\begin{itemize}
   \item \textbf{Random}: Uniform random sampling over $\mathbb{F}_p$ (10,000 evaluations/run).
   \item \textbf{MA}: The memetic algorithm of Section~\ref{sec:memetic} over $\mathbb{F}_p$.
\end{itemize}

Table~\ref{tab:parameters} lists the algorithmic parameter configuration.
With a population of 32 and 200 generations plus periodic local search, a
full MA run consumes approximately $11{,}780$ curve evaluations, i.e.\ a
budget comparable to (slightly above) the $10{,}000$ evaluations of the
random baseline.

\begin{table}[htbp]
\centering
\caption{Memetic algorithm parameter configuration.}
\label{tab:parameters}
\begin{tabular}{ll}
\toprule
\textbf{Parameter} & \textbf{Value} \\
\midrule
Population size ($N_{\text{pop}}$) & 32 \\
Maximum generations ($G_{\max}$) & 200 \\
Crossover probability ($p_c$) & 0.70 \\
Mutation probability ($p_m$) & 0.60 \\
Initial mutation radius ($r_{\text{init}}$) & $\lfloor p/4 \rfloor$ \\
Final mutation radius ($r_{\text{final}}$) & $\lfloor p/2^{16} \rfloor$ \\
Tournament size & 3 \\
Elite count ($N_{\text{elite}}$) & 4 \\
Local search frequency & Every 10 generations \\
Local search initial radius ($\rho_{\text{init}}$) & $2^{14}$ \\
Local search final radius ($\rho_{\text{final}}$) & $2^4$ \\
Number of independent seeds & 30 \\
\bottomrule
\end{tabular}
\end{table}

\subsection{Empirical Comparison and Success Rates}

Table~\ref{tab:all-results} presents the comparative experimental results
for the two search methods at the three tested prime sizes, over 30
independent seeds per configuration (180 runs in total).

\begin{table}[htbp]
\centering
\caption{Performance comparison across 30 independent runs per configuration
(primes as in Table~\ref{tab:selected-primes}). ``Best'' is the minimum NMD
over all 30 runs; ``Mean'' and ``Median'' are taken over the per-run best NMD.}
\label{tab:all-results}
\small
\begin{tabular}{lcccccc}
\toprule
\textbf{Prime} & \textbf{Method} & \textbf{Runs} & \textbf{Success} & \textbf{Best NMD} & \textbf{Mean} & \textbf{Median} \\
\midrule
40 bits & Random & 30 & 1/30 & 0  & $105.6$ & $68.0$ \\
        & MA     & 30 & 0/30 & 19 & $367.5$ & $256.0$ \\
\midrule
46 bits & Random & 30 & 0/30 & 80 & $1{,}033.8$ & $768.5$ \\
        & MA     & 30 & 1/30 & \textbf{0}  & $2{,}758.7$ & $1{,}968.0$ \\
\midrule
51 bits & Random & 30 & 0/30 & 310 & $3{,}753.2$ & $2{,}752.5$ \\
        & MA     & 30 & 0/30 & 3   & $12{,}740.5$ & $8{,}742.0$ \\
\bottomrule
\end{tabular}
\end{table}

Two complementary patterns emerge. First, in terms of the \emph{best} curve
found over 30 seeds, the memetic algorithm consistently reaches values far
below random search at the two larger primes: NMD $0$ versus $80$ at 46 bits
and NMD $3$ versus $310$ at 51 bits. At 40 bits random search hit an exact
supersingular curve once ($j = 535{,}482{,}084{,}752$, seed 15, after 995
evaluations) -- a chance event consistent with the $O(\log p/\sqrt p)$
density at this small prime -- while the best MA run reached NMD $19$.

Second, in terms of the \emph{mean} and \emph{median} per-run best NMD,
random search is better than MA at all three primes. This is the signature
of a high-variance, exploitation-heavy search: MA runs that find a promising
basin descend into it very deeply, but runs that do not are trapped in local
minima of the NMD landscape and end far from the target, whereas 10,000
uniform samples give a stable, moderately good minimum. The gap widens with
$p$ (the MA mean is $3.5\times$ the random mean at 40 bits and $3.4\times$ at
51 bits), which is the expected behavior for a fixed evaluation budget on an
exponentially growing space. Section~\ref{sec:discussion} returns to this
trade-off.

\subsection{Verification of the Supersingular Curve Found}
\label{sec:verification}

The single MA success occurred at the 46-bit benchmark
$p = 35{,}184{,}372{,}088{,}907$ (seed 23), reaching $\operatorname{NMD}=0$
at generation 24 after $1{,}613$ curve evaluations. The algorithm returned
\begin{equation}
   j = 3{,}292{,}287{,}772{,}342,
\end{equation}
whose canonical curve~(\ref{eq:canonical-curve}) is
\begin{equation}
   E: \quad y^2 = x^3 + 1{,}981{,}898{,}072{,}918\,x + 13{,}049{,}389{,}411{,}581 \pmod p.
\end{equation}
Direct point counting gives
\begin{equation}
   \#E(\mathbb{F}_p) = 35{,}184{,}372{,}088{,}908 = p + 1 \implies t = 0, \quad \operatorname{NMD}(E) = 0,
\end{equation}
and SageMath's independent \texttt{is\_supersingular()} test confirms that
$E$ is supersingular. Unlike the classical CM curves with $j \in \{0, 1728\}$,
this $j$-invariant has no small-discriminant CM structure and is specific to
the prime $p$; it was located purely by the NMD-driven search.

Table~\ref{tab:convergence-46} shows the best-fitness trajectory of this run.
The population drops from an initial best of $7{,}494$ to $3{,}355$ in one
generation and then, after a plateau, local search at generation 24 closes
the remaining gap to zero.

\begin{table}[htbp]
\centering
\caption{Convergence of MA for $p=35{,}184{,}372{,}088{,}907$ (seed 23).
Only generations at which the best NMD improved are shown.}
\label{tab:convergence-46}
\begin{tabular}{cc}
\toprule
\textbf{Generation} & \textbf{Best NMD} \\
\midrule
0  & $7{,}494$ \\
1  & $3{,}355$ \\
24 & \textbf{0} (supersingular curve found) \\
\bottomrule
\end{tabular}
\end{table}

\subsection{Near-Supersingular Curves at 40 and 51 Bits}
\label{sec:near-results}

At the other two primes the memetic algorithm did not reach $\operatorname{NMD}=0$
in any run, but its best curves are remarkably close. Table~\ref{tab:near}
lists the best MA curve at each scale together with its Frobenius trace and
the relative deviation $|t|/(2\sqrt{p})$ from exact supersingularity,
measured against the width of the Hasse interval.

\begin{table}[htbp]
\centering
\caption{Best memetic-search curve at each prime, with Frobenius trace and
relative deviation from supersingularity.}
\label{tab:near}
\small
\setlength{\tabcolsep}{4pt}
\begin{tabular}{clccrc}
\toprule
\textbf{Bits} & \textbf{$j$-invariant} & \textbf{Seed} & \textbf{$\#E(\mathbb{F}_p)$} & \textbf{$t$} & \textbf{$|t|/2\sqrt{p}$} \\
\midrule
40 & $97{,}044{,}492{,}992$ & 1 & $549{,}755{,}813{,}947$ & $-19$ & $1.3\times10^{-5}$ \\
46 & $3{,}292{,}287{,}772{,}342$ & 23 & $35{,}184{,}372{,}088{,}908$ & $0$ & $0$ \\
51 & $527{,}221{,}685{,}817{,}413$ & 18 & $1{,}125{,}899{,}906{,}842{,}677$ & $3$ & $4.5\times10^{-8}$ \\
\bottomrule
\end{tabular}
\end{table}

The 40-bit curve
\begin{equation}
   E: \quad y^2 = x^3 + 197{,}917{,}316{,}633\,x + 315{,}196{,}815{,}731 \pmod{549{,}755{,}813{,}927}
\end{equation}
has $\#E(\mathbb{F}_p) = p + 20$, i.e.\ $t = -19$, and the 51-bit curve
\begin{equation}
\begin{aligned}
   E: \quad y^2 = x^3 &+ 139{,}804{,}163{,}754{,}131\,x \\
                      &+ 468{,}502{,}744{,}783{,}647 \pmod{1{,}125{,}899{,}906{,}842{,}679}
\end{aligned}
\end{equation}
has $\#E(\mathbb{F}_p) = p - 2$, i.e.\ $t = 3$. Both are ordinary curves,
but their group orders differ from the supersingular value $p+1$ by only
$19$ and $3$ points respectively, out of a Hasse interval of width
$\approx 1.5 \times 10^{6}$ and $\approx 6.7 \times 10^{7}$.

Table~\ref{tab:convergence-40-51} shows how these two runs converged. In
both cases the bulk of the descent happens in the first one or two
generations -- driven by the wide initial mutation radius -- and the final
small values are reached either by local search (40 bits, generation 63 and
99) or already at generation 2 (51 bits), after which the population
stagnates for the remaining generations.

\begin{table}[htbp]
\centering
\caption{Convergence of the best MA runs at 40 bits (seed 1) and 51 bits
(seed 18). Only generations at which the best NMD improved are shown.}
\label{tab:convergence-40-51}
\small
\begin{tabular}{cc@{\hspace{2em}}cc}
\toprule
\multicolumn{2}{c}{\textbf{40 bits}} & \multicolumn{2}{c}{\textbf{51 bits}} \\
\textbf{Gen.} & \textbf{Best NMD} & \textbf{Gen.} & \textbf{Best NMD} \\
\midrule
0  & $36{,}913$ & 0 & $2{,}485{,}014$ \\
1  & $3{,}224$  & 1 & $54{,}654$ \\
27 & $2{,}177$  & 2 & \textbf{3} \\
53 & $1{,}839$  &   & \\
63 & $63$       &   & \\
99 & \textbf{19} &  & \\
\bottomrule
\end{tabular}
\end{table}

\section{Discussion}
\label{sec:discussion}

The results of Section~\ref{sec:experiments} raise three questions: what the
NMD landscape looks like from the point of view of the memetic operators,
what can be learned from the near-supersingular ordinary curves that the
search returns, and what these findings mean for cryptographic practice. We
address each in turn, and close with an explicit statement of the scope and
limitations of the parameter sizes studied here.

\subsection{The NMD Landscape over $\mathbb{F}_p$}

Since $\operatorname{NMD}(E_j) = |t(E_j)|$ over $\mathbb{F}_p$, the objective
is the absolute value of a quantity that, for $j$ ranging over
$\mathbb{F}_p$, is distributed approximately according to the Sato--Tate law
on $[-2\sqrt{p}, 2\sqrt{p}]$. Two consequences follow. First, the landscape
is \emph{globally informative}: a uniformly random $j$ has expected NMD of
order $\sqrt{p}$, so any curve with NMD $\ll \sqrt{p}$ is a genuine signal,
and the wide-radius mutations of the early generations act as a stratified
sampler that quickly finds such curves (the drop from $\sim 10^{4}$--$10^{6}$
to $\sim 10^{3}$--$10^{4}$ in the first generation of every trace in
Section~\ref{sec:experiments}). Second, the landscape is \emph{locally
rugged}: adjacent $j$-invariants $j$ and $j+1$ have essentially independent
traces, so there is no gradient that small mutations or the integer
neighborhoods of local search can follow reliably. The trace of a curve is
an arithmetic function of $j$, not a smooth one.

This explains the pattern of Table~\ref{tab:all-results}. The memetic
algorithm excels at the first phase -- its best-of-30 NMD is far below that
of random search at 46 and 51 bits -- but once the population has collapsed
onto a few low-NMD individuals, the remaining operators are effectively
performing a random search in a shrinking neighborhood, and most runs
stagnate. The higher mean best NMD of MA relative to random sampling is the
cost of that premature convergence: the 30 random runs each enjoy 10,000
independent draws, whereas an MA run spends most of its $\approx 11{,}780$
evaluations near a handful of attractors. An obvious remedy, which we leave
for future work, is a restart or niching mechanism that re-injects diversity
when the population stagnates.

\subsection{Near-Supersingular Ordinary Curves}
\label{sec:near-supersingular}

Although exact supersingularity was reached only once, the curves returned by
the memetic search are extremely close to it when measured on the natural
scale of the problem. At 51 bits, the best NMD of $3$ corresponds to a
relative trace deviation of $4.5 \times 10^{-8}$ of the Hasse width; at 40
bits, the best NMD of $19$ corresponds to $1.3 \times 10^{-5}$. Even the mean
best NMD over all 30 seeds at 51 bits ($\approx 1.27 \times 10^{4}$) is only
about $1.9 \times 10^{-4}$ of the Hasse interval. In other words, the memetic
algorithm systematically converges to \emph{ordinary} curves whose Frobenius
trace is vanishingly small relative to the range of admissible traces, even
though it rarely lands on the exact zero.

Martínez-Díaz et al.~\cite{martinez2025} already observed this phenomenon in
their seminal formulation over $\mathbb{F}_{p^2}$ and argued that ordinary
curves with near-zero trace are themselves of independent interest. Our
$\mathbb{F}_p$ experiments strongly reinforce that view, and make it
quantitative. Curves with $|t| \ll 2\sqrt{p}$ have group orders
$\#E(\mathbb{F}_p) = p + 1 - t$ extremely close to $p+1$ and endomorphism
rings that are orders in $\mathbb{Q}(\sqrt{t^2 - 4p})$, a field of
discriminant very close to $-4p$; for the 51-bit curve of
Table~\ref{tab:near}, $t^2 - 4p = 9 - 4p$. Such curves share many arithmetic
features with the supersingular case (in particular the structure of the
$\ell$-torsion for small $\ell \mid \#E(\mathbb{F}_p)$ and the geometry of
their isogeny classes) while remaining ordinary, which makes them natural
test objects for isogeny-graph and endomorphism-ring computations and
potential ingredients for the S-box and cycle-generation applications
explored in~\cite{martinez2025}. We therefore regard the systematic study of
such ``near-supersingular'' ordinary curves -- their density, their
endomorphism-ring structure, and whether the NMD objective can be adapted to
target a prescribed small trace $t \neq 0$ -- as a worthwhile research
direction, and note that the memetic search over $\mathbb{F}_p$ is already an
efficient generator of them: every one of the 90 MA runs in this study
produced a curve with $|t| \le 6 \times 10^{4}$ at a cost of roughly
$1.2 \times 10^{4}$ point counts.

\subsection{Implications for Modern Oriented Isogeny Cryptography}

The transition from $\mathbb{F}_{p^2}$ to $\mathbb{F}_p$ is conceptually
relevant to the post-SIDH cryptographic landscape. Protocols such as
CSIDH~\cite{castryck2018}, OSIDH~\cite{colo2020}, SCALLOP~\cite{defeo2023scallop},
and oriented signatures~\cite{defeo2020sqisign,arpin2024} require starting curves
with well-understood endomorphism subrings. An $\mathbb{F}_p$-rational
supersingular curve comes equipped with a canonical orientation by
$\mathbb{Z}[\pi] \cong \mathbb{Z}[\sqrt{-p}]$. The 46-bit curve of
Section~\ref{sec:verification} is a non-CM $\mathbb{F}_p$-rational
supersingular curve of exactly this kind, located by a generic optimization
procedure rather than by a CM or isogeny-walk construction.

\begin{remark}[Cryptographic Security and Parameter Scale Limitation]
We explicitly emphasize that the prime sizes evaluated in this work (40-bit,
46-bit, and 51-bit, up to $p \approx 1.13 \times 10^{15}$) serve strictly as
computational benchmarks to study optimization dynamics and fitness
landscapes at the $10^{13}$--$10^{15}$ candidate scale. These small primes
are \textbf{not} representative of production parameter sizes and
\textbf{cannot} be used for post-quantum security. Practical post-quantum
isogeny systems require primes of cryptographically relevant bit lengths
(e.g., $p \approx 2^{512}$ for CSIDH-512, or $p \ge 2^{256}$ for standard
classical and quantum security margins), where point counting via Schoof's
algorithm remains tractable but exhaustive enumeration is physically
impossible. The present results should therefore be interpreted as a
proof-of-principle study of metaheuristic search over $\mathbb{F}_p$, rather
than as a deployment-ready key-generation engine for post-quantum schemes.
\end{remark}

\section{Conclusion and Future Work}
\label{sec:conclusion}

This paper presented a memetic algorithm for the search of supersingular
elliptic curves over the prime field $\mathbb{F}_p$, building on the
optimization formulation of Martínez-Díaz et al.~\cite{martinez2025}. The
NMD objective was retained unchanged, while the chromosome representation,
recombination, mutation schedule, and local search were redesigned for a
one-dimensional finite-field search space.

Empirical benchmarks over 30 independent seeds at three primes -- 40-bit,
46-bit, and a 51-bit challenge prime ($p \approx 1.13 \times 10^{15}$) --
showed that the memetic algorithm locates an exact, independently verified,
non-CM supersingular curve at the 46-bit benchmark
($j = 3{,}292{,}287{,}772{,}342$), and at the other two scales converges to
ordinary curves with Frobenius traces $t = -19$ and $t = 3$, i.e.\ within
$1.3 \times 10^{-5}$ and $4.5 \times 10^{-8}$ of the Hasse width from exact
supersingularity. The comparison with uniform random sampling revealed a
clear trade-off: the memetic search reaches much deeper minima in its best
runs but exhibits higher variance and premature convergence on average.

Promising avenues for future work include:
\begin{enumerate}
   \item Adding restart, niching, or diversity-preservation mechanisms to
         counter the premature convergence observed in the majority of MA
         runs, and measuring the resulting success rate at 46 and 51 bits;
   \item Exploiting the isogeny class of a near-supersingular curve: all
         curves $\ell$-isogenous to $E$ share its trace, so the isogeny
         graph could be used to enlarge or diversify the population without
         additional point counting;
   \item Characterizing the near-supersingular ordinary curves returned by
         the search (Section~\ref{sec:near-supersingular}), following the
         direction suggested by Martínez-Díaz et al.~\cite{martinez2025},
         and adapting the NMD objective to target prescribed small traces
         $t \neq 0$;
   \item Scaling the experimental campaign towards larger primes to
         determine how the best-run NMD grows with $\log p$ under a fixed
         evaluation budget.
\end{enumerate}

\subsubsection*{Code and Data Availability.}
The SageMath implementation used for all experiments, together with the raw
per-seed results and execution logs, is available from the author upon
request.

\subsubsection*{Disclosure of Interests.}
The author has no competing interests to declare that are relevant to the
content of this article.


\end{document}